# Improved energy storage and electrocaloric properties of lead-free $Ba_{0.85}Ca_{0.15}Zr_{0.1}Ti_{0.9}O_3$ ceramic

Afaak Lakouader[1]*, Hanane Mezzourh[1, 2], Daoud Mezzane[1, 2], M'barek Amjoud[1], Lahoucine Hajji[1], El Hassan Choukri[1], Igor A. Luk'yanchuk [2,3], Zdravko Kutnjak[4] , and Mimoun El Marssi[2]

[1] IMED-Lab, Cadi Ayyad University, Marrakesh, 40000, Morocco

[2] LPMC, University of Picardy Jules Verne, Amiens, 80039, France

[3] Department of Building Materials, Kyiv National University of Construction and Architecture, Kyiv, 03680, Ukraine

[4] Jozef Stefan Institute, Jamova Cesta 39, 1000 Ljubljana, Slovenia

* Corresponding author: afaak.lakouader@gmail.com

**Abstract**

Lead-free $Ba_{0.85}Ca_{0.15}Zr_{0.1}Ti_{0.9}O_3$ (BCZT) ceramic powders were synthesized using the sol-gel method. To achieve high-energy storage and large electrocaloric effect in bulk ceramics the ceramics thickness was reduced. Dielectric, ferroelectric, energy storage, and electrocaloric properties were investigated for BCZT ceramic with 400 µm. Here, pure crystalline structure and homogenous microstructure were identified by XRD analysis and SEM measurements, respectively. The dielectric measurements revealed a maximum dielectric constant associated with ferroelectric-paraelectric phase transition. The maximum of $\varepsilon_r'$ was 7841 around 352 K. Furthermore, the BCZT ceramic showed improved energy storage and electrocaloric properties. A high recoverable energy density $W_{rec}$ of 0.24 J/cm$^3$ and a total energy density $W_{total}$ of 0.27 J/cm$^3$ with an efficiency coefficient of ~ 88% at 423 K under an electric field of 55 kV/cm was obtained. Besides, The maximum value of $\Delta T$ = 2.32 K, the electrocaloric responsivity $\zeta$ = 0.42 K mm/kV, the refrigeration capacity $RC$= 4.59 J/kg and the coefficient of performance $COP$ = 12.38 were achieved around 384 K under 55 kV/cm. The total energy density $W_{total}$ and the temperature change $\Delta T$ were also calculated by exploiting the Landau-Ginzburg-Devonshire (LGD) theory. The theoretical results matched the experimental findings. These results suggest that the synthesized BCZT ceramic with the reduced thickness could be a promising candidate for energy storage and electrocaloric applications.

**Declarations**

**Funding**

CNRST PriorityProgram PPR 15/2015;

The European Union's Horizon 2020 research ;

The Ministry of education and science of the Russian Federation Project #13.2251.21.0042.

**Conflicts of interest/Competing interests**

Not applicable

**Availability of data and material**

Not applicable

**Code availability**

Not applicable

**Ethics approval**

Not applicable

**Consent to participate**

We confirm that all authors mentioned in the manuscript have participated in, read and approved the manuscript, and have given their consent for the submission and subsequent publication of the manuscript.

**Consent for publication**

We confirm that all the authors mentioned in the manuscript have agreed to publish this paper.

**Relevance Summary**

In this study, we report successful preparation of lead-free $Ba_{0.85}Ca_{0.15}Zr_{0.1}Ti_{0.9}O_3$ ceramic with reduced thickness in order to reach high electric field. The study of their structural, dielectric, ferroelectric were also reported. The energy storage and electrocaloric properties were investigated by using the indirect experimental approach based on the Maxwell relation and the Landau-Ginzburg-Devonshire (LGD) theory. We confirm that this work is original and has not been published elsewhere nor it is not currently under consideration for publication elsewhere.

**Acknowledgments**

The authors gratefully acknowledge the generous financial support of CNRST Priority Program PPR 15/2015, the European Union's Horizon 2020 research and the Ministry of education and science of the Russian Federation Project #13.2251.21.0042.

**Introduction**

Ferroelectric materials can be defined as dielectric materials in which permanent polarization exists even after removing the applied electric field. For most applications of ferroelectric materials, the dielectric constant and loss are essential practical parameters. These applications include non-volatile memories, pyroelectric detectors, sensors, actuators, transducers, capacitors, generators, optical memories, and energy harvesting devices [1].

The most used ferroelectric materials are oxides with a perovskite structure of $ABO_3$. A cubic lattice of space group (*Pm-3m*) describes this high symmetry structure [2]. Where A is an alkali cation, alkaline earth, or rare earth, B is a transition cation of atomic radius smaller than that of site A. The structure is a network of corner-linked oxygen octahedral, with the smaller cation filling the octahedral holes and the large cation filling the dodecahedral holes.

Among perovskite ferroelectric materials, the most studied are lead titanate ($PbTiO_3$) [3], lead zirconate titanate (PZT) [4], lanthanum zirconate titanate (PLZT) [5]. However, lead is recognized as toxic, which is dangerous to humans and the environment. Therefore, to replace lead-based materials, it is necessary to develop environmentally friendly materials with no lead [6]. Thus, several scientific works have focused mainly on $BaTiO_3$ (BT) perovskite because it has interesting advantages such as low dielectric loss and stable electrical properties for long-term operations [7]. However, their use is limited due to their relatively high curie temperature *Tc* (~390 K) [8]. Therefore, adequate atomic substitutions are expected to modify the chemical bonds of the structure and, thus, the mobility of the ions and the resulting polarizability. A series of studies have been initiated to modify the dielectric and ferroelectric properties of $BaTiO_3$ systems, including $(Ba,Ca)TiO_3$ [9], $Ba(Ti,Zr)O_3$ [10], $(Ba,Ca)(Ti,Sn)O_3$ [11],etc. In these doped (BT) matrix systems, the dielectric loss of $Ba(Ti,Zr)O_3$(BZT) ceramics is very low, which can be attributed to the increased substitution rate of zirconium [12]. In addition, the good temperature stability of (Ba, Ca) $TiO_3$ (BCT) ceramics is due to the introduction of Ca at site A of the BT system, which leads to the diffuse phase transition [13].

More recently, $(Ba,Ca)(Zr,Ti)O_3$ (BCZT) materials with $Ca^{2+}$ and $Zr^{4+}$ at site A and site B of the $BaTiO_3$ perovskite structure have been widely studied for their excellent dielectric, ferroelectric and piezoelectric properties, and their composition closed to the morphotropic phase boundary MPB [14]. Among these ceramics, xBa $(Zr_{0.2}Ti_{0.8})$ $O_3$– (1–x) $(Ba_{0.7}Ca_{0.3})$ TiO3 ceramics with (0.48 <x <0.52) were largely studied, and the results showed that the composition $Ba_{0.85}Ca_{0.15}Zr_{0.1}Ti_{0.9}O_3$ has considerable dielectric, ferroelectric and piezoelectric properties compared to other components [15]. In this regard, Wang et al. reported that $Ba_{0.85}Ca_{0.15}Zr_{0.1}Ti_{0.9}O_3$ ceramics exhibit a high spontaneous polarization ($P_s$=9.69 μC/cm$^2$), a low coercive field ($E_c$ =1.89 kV/cm) and a high dielectric permittivity ($\varepsilon_r'$=16310) [16]. In addition, due to low remnant polarization, high maximal polarization, and adjustable breakdown strength (BDS), relaxor ferroelectrics (RFEs), including BCZT, have a promising future in the field of dielectric energy, and their electric polarization response can be converted into recoverable electrostatic energy.

One of the crucial aspects of functional microstructure is the synthesis method. Many papers have reported that the sol-gel process should be a rather promising approach as it offers the possibility of mixing metal ions at the atomic level [17], [18]. Besides, this method also has several potential advantages of better homogeneity, chemical purity, and stoichiometric composition, which can be difficult to achieve by the solid-state method [19]. With a variation of the sintering temperature, various BCZT properties can be extracted. X. Ji et al. reported that BCZT sintered at 1400 °C had a uniform microstructure with a density of 5.56 g/cm$^3$ and had excellent electrical properties ($\varepsilon_r'$ = 8808, $P_r$ =12.24 µC /cm$^2$, $d_{33}$ = 485 pC/N) [20]. Moreover, BCZT ceramics obtained by sol-gel method and sintered at 1420 °C have outstanding electrical and energy storage properties ($d_{33}$ = 558 pC/N, $\varepsilon_r'$ = 3375, $tg\delta$ = 0.02, and $W$ = 0.52 J/cm$^3$) as evaluated by Zhong Ming Wang et al. [17]. The energy density can also be demonstrated by exploiting the Landau-Ginzburg-Devonshire (LGD) phenomenological theory. The modelling result supports the experimental findings as reported for $BaTiO_3$-$BaSnO_3$ by Yonggang et al. [21]. Meanwhile, a large electrocaloric responsivity ($\xi = \Delta T / \Delta E$) was found ($\xi$ = 0.19 K mm/kV) for BCZT ceramic sintered at 1420 ° C as reported by H. Kaddoussi et al. [22]. Additional estimations of the EC effect in ferroelectric materials can be obtained by theoretical investigations exploiting the LGD phenomenological approach.

A different approach has been used to improve the electrocaloric efficiency (1) by monitoring the perovskite composition, (2) by combining ferroelectric, relaxor ferroelectric, and antiferroelectric materials, (3) by using composites (ceramics/polymers), and (4) by reducing the ceramic thickness. On the other hand, according to previous studies on the energy storage density of material dielectrics, a higher breakdown strength is responsible for a larger energy storage density [23]. In this regard, the breakdown strength of dielectrics increases exponentially with decreasing ceramic's thickness [24]. Thus, reducing thickness is a practical method to enhance the breakdown strength and energy density for dielectrics. The dielectric thin films exhibit high breakdown strength, large energy storage density and giant electrocaloric effect compared to bulk ceramics. However, their energy capacities are low because of the low thickness or volume of thin films, making them unsuitable for cooling applications where a high level of energy is needed. This indicates that thin bulk ceramics should be more appropriate for practical energy storage density and electrocaloric effect. Thus, to achieve higher energy storage properties and a giant electrocaloric effect in bulk ceramics, enhancing the breakdown strength is necessary to tolerate strong electric fields while keeping a large dielectric volume. This can be accomplished by decreasing the ceramic's thickness.

In this work, we investigated the temperature dependency of energy storage and electrocaloric properties of BCZT ceramic prepared by sol–gel using the indirect experimental approach based on the Maxwell relation and the Landau-Ginzburg-Devonshire (LGD) theory.

## Experimental

The $Ba_{0.85}Ca_{0.15}Zr_{0.1}Ti_{0.9}O_3$ ceramic were synthesized by the sol-gel route. Stoichiometric amounts of barium acetate $Ba(CH_3COO)_2$(99%), calcium nitrate tetrahydrate $Ca(NO_3)_2.4H_2O$ (99%), zirconyl chloride octahydrate $ZrOCl_2.8H_2O$ (98%), and titanium isopropylate $C_{12}H_{28}O_4Ti$ (97%) were used as precursors. As solvents, acetic acid (100%) and 2-Methoxyethanol (99%) were chosen. $Ba(CH_3COO)_2$ and $Ca(NO_3)_2.4H_2O$ were initially dissolved in acetic acid, while [$ZrOCl_2.8H_2O$] was dissolved in 2-Methoxyethanol, then an appropriate amount of $C_{12}H_{28}O_4Ti$ was added. The precursor solution of (Zr,Ti) was added to that of (Ba,Ca) and stirred for 1 h at room temperature to give a BCZT solution, and then the pH was adjusted with ammonia to 7. The sol obtained was dried in air for 48 hours to remove the solvents and then ground into a fine powder in a mortar. The powder was calcined for 4 h at 1000 ° C in the air to form a pure crystalline phase. The calcined powder was then pressed into cylindrical discs by uniaxial pressing to prepare pellets sintered at a high temperature of 1420 ° C for 4 h to improve the ceramic's density. The pellets were polished to adjust the ceramic thickness at 400 µm.

The sintered ceramic's phase structure was determined by X-ray diffraction (XRD, Panalytical X-Pert Pro) analysis using the Cu-Kα radiation with λ~1.540598 Å. The measurements were performed at room temperature in a range of 2θ from 20° to 80° with a step size of 0.02°. Rietveld refinement analysis was realized by using FullProf software. A Scanning Electron Microscope (SEM, VEGA 3-Tscan) revealed the surface's morphologies. The density of the pellet was measured using Archimedes' method. The dielectric measurements were examined using an HP 4284A precision impedance meter, controlled by a computer in the 20Hz to 1MHz frequency range. The surfaces of the sintered sample were covered with a conductive silver paste serving as electrodes for electrical measurements. The polarization-electric field (P-E) hysteresis loops were collected using CPE1701, PolyK, USA, with a high voltage power supply (Trek 609-6, USA). The electrocaloric study was achieved by the indirect method from recorded (P-E) hysteresis loops at 10 Hz as a function of temperature.

## Results and discussion

### I. Structural and microstructural properties

**Fig. 1** presents the X-ray diffraction pattern of BCZT ceramic sintered at 1420°C for 4h. The formation of a complete single-phase perovskite structure, with any impurity phase, is evident in the room temperature XRD data, suggesting that $Ca^{2+}$ and $Zr^{4+}$ have been diffused into the $BaTiO_3$ lattice to form a solid solution [25]. The peak between $2\theta \approx 45°$ was split into two peaks corresponding to (002) and (200) planes, as represented in **Fig. 1b**. This could be the reflection of a tetragonal structure [26]. The splitting peaks of (022) and (200) around $2\theta \approx 45°$ imply the existence of an orthorhombic phase which can also be confirmed by the formation of the triplet around $2\theta \approx 66°$ [27], as displayed in **Fig. 1c**. The coexistence of orthorhombic and tetragonal phases indicates the formation of BCZT ceramic at the morphotropic phase boundary (MPB). The coexistence of tetragonal and orthorhombic phases at room temperature was also reported in many reports [16], [28]. The data of XRD was analyzed by Rietveld refinement, which was achieved by using a combination of tetragonal symmetry with space group *P4mm* and orthorhombic symmetry with space group *Pmm2* ($\chi^2$=1.86, *Rp*=17.1 and *Rwp*=13.5). Structural information and phase composition are regrouped in Table 1.

**Fig. 2** shows the surface morphology of the sample of the BCZT ceramic sintered at 1420 °C. The grains of the ceramic show an irregular morphology and a uniform distribution. The average grain size is 45 µm. The SEM shows that the ceramic displays a dense morphology due to the high sintering temperature. The latter facilitates the mass transfer and pore removal resulting in tightly packed grains and a high density of 5.70 g/cm$^3$, which is ~ 95% of the theoretical density.

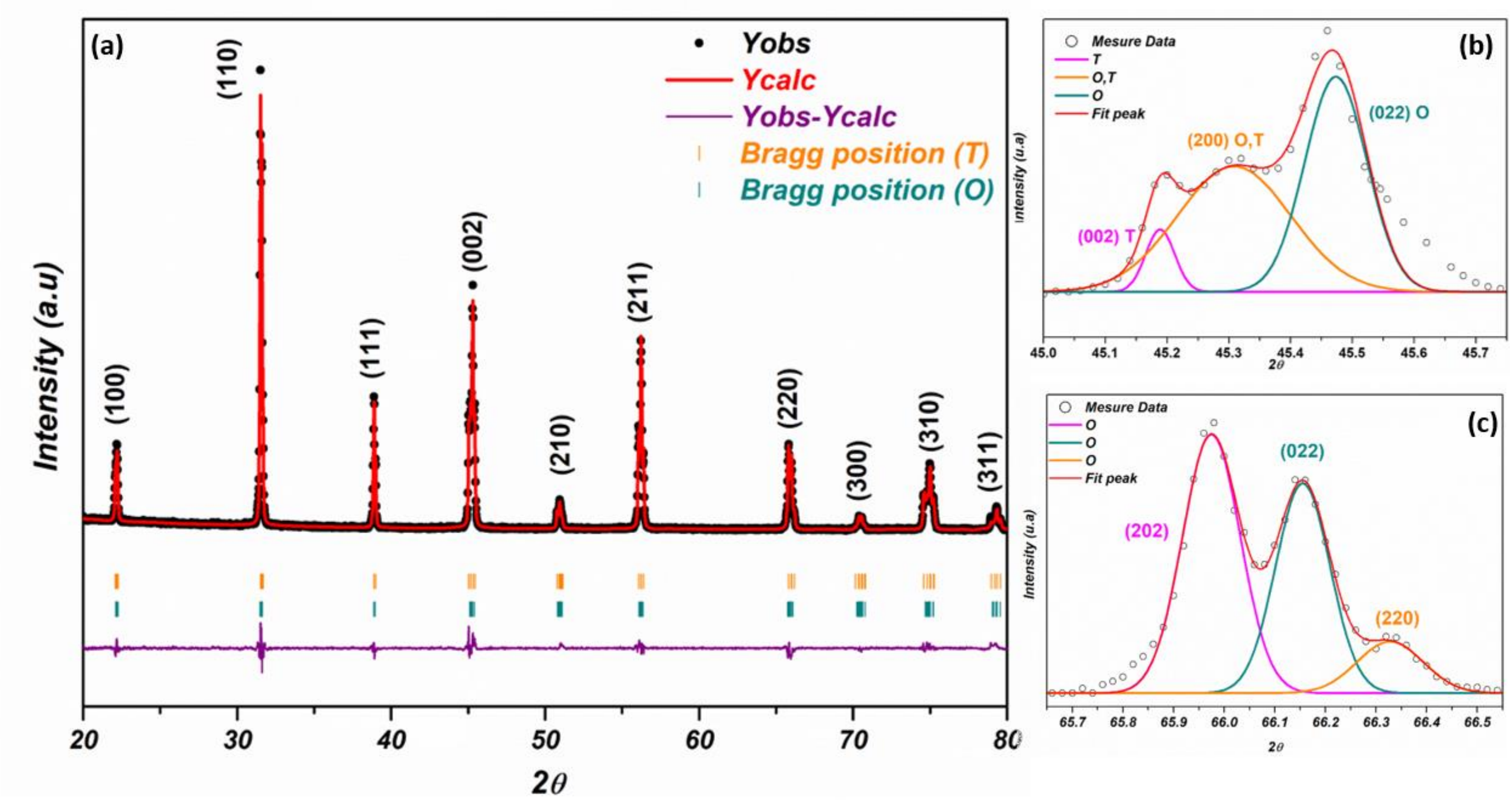

Figure 1: (a) XRD patterns and Rietveld refinement of BCZT sample sintered at 1420°C for 4h; (b) XRD patterns for 2Θ between [45°- 46°] (c) XRD patterns for 2Θ between [65 °-67°]

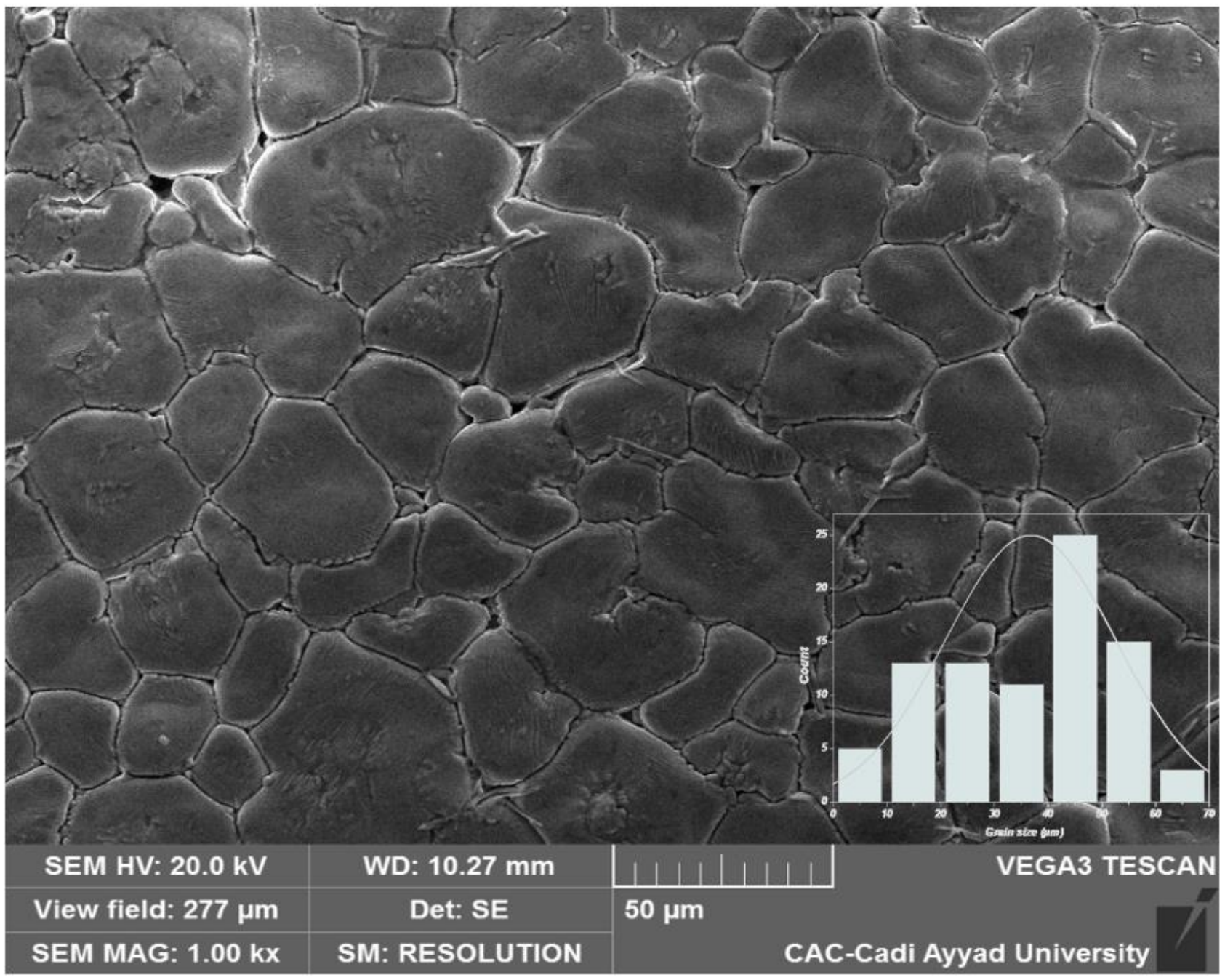

Figure 2: The SEM image of sintered BCZT ceramic sintered at 1420°C

Table1: Structural parameters, average crystalline size, and relative density of BCZT ceramic

| Structure | | Tetragonal | Orthorhombic |
|---|---|---|---|
| Space Group | | *P4mm* | *Pmm2* |
| unit cell parameters | $a$(Å) | 3.999900 | 4.001126 |
| | $b$(Å) | 3.999900 | 4.016704 |
| | $c$(Å) | 4.022146 | 4.011608 |
| | $\alpha=\beta=\delta$ | 90° | 90° |
| *Volume* ($Å^3$) | | 64.35 | 64.47 |
| ($c/a$) | | 1.005 | |
| Phase composition (wt %) | | 66.51 | 33.49 |
| Reliability factors (%) | $\chi 2$ | 1.86 | |
| | *Rp* | 17.1 | |
| | *Rwp* | 13.5 | |
| Average grain size (µm) by SEM | | 45 | |
| Relative density (%) | | 95 | |

## II. Dielectric properties

**Fig. 3** shows the thermal variations of the real part of the dielectric permittivity ($\varepsilon_r$') as a function of the frequency for BCZT ceramic; the inset shows the dielectric losses (*tgδ*). The results show that $\varepsilon_r$' increases rapidly then decreases with increasing measurement temperature. This behavior is the same as a conventional ferroelectric material [29]. Tow dielectric anomalies, $T_{O\text{-}T}$ and $T_{C,}$ were observed around 300 and 352 K, corresponding to the phase transition from orthorhombic to tetragonal and from tetragonal to cubic, respectively. The response of space charge polarization induced by the creation of lattice vacancies, such as cation or oxygen vacancies, could cause a high $\varepsilon_r$' at a low frequency [26]. However, as frequency increase, space charge polarizations became more limited, resulting in a decrease in $\varepsilon_r$' [26]. At 1 kHz, the maximum of *εr'* were 17841 around 352 K. The dielectric loss *tgδ* was less than 0.06, which may be attributed to the dense microstructure and the lower electron diffusion in the grain boundaries [33]. Compared to BCZT synthesis by solid state and sintered at 1350 °C, the maximum of $\varepsilon_r$' was 6330 around 388 K [30], which is lower than our results. The increase in $\varepsilon_r$' could be caused by increasing density, which is essential for the electrocaloric effect.

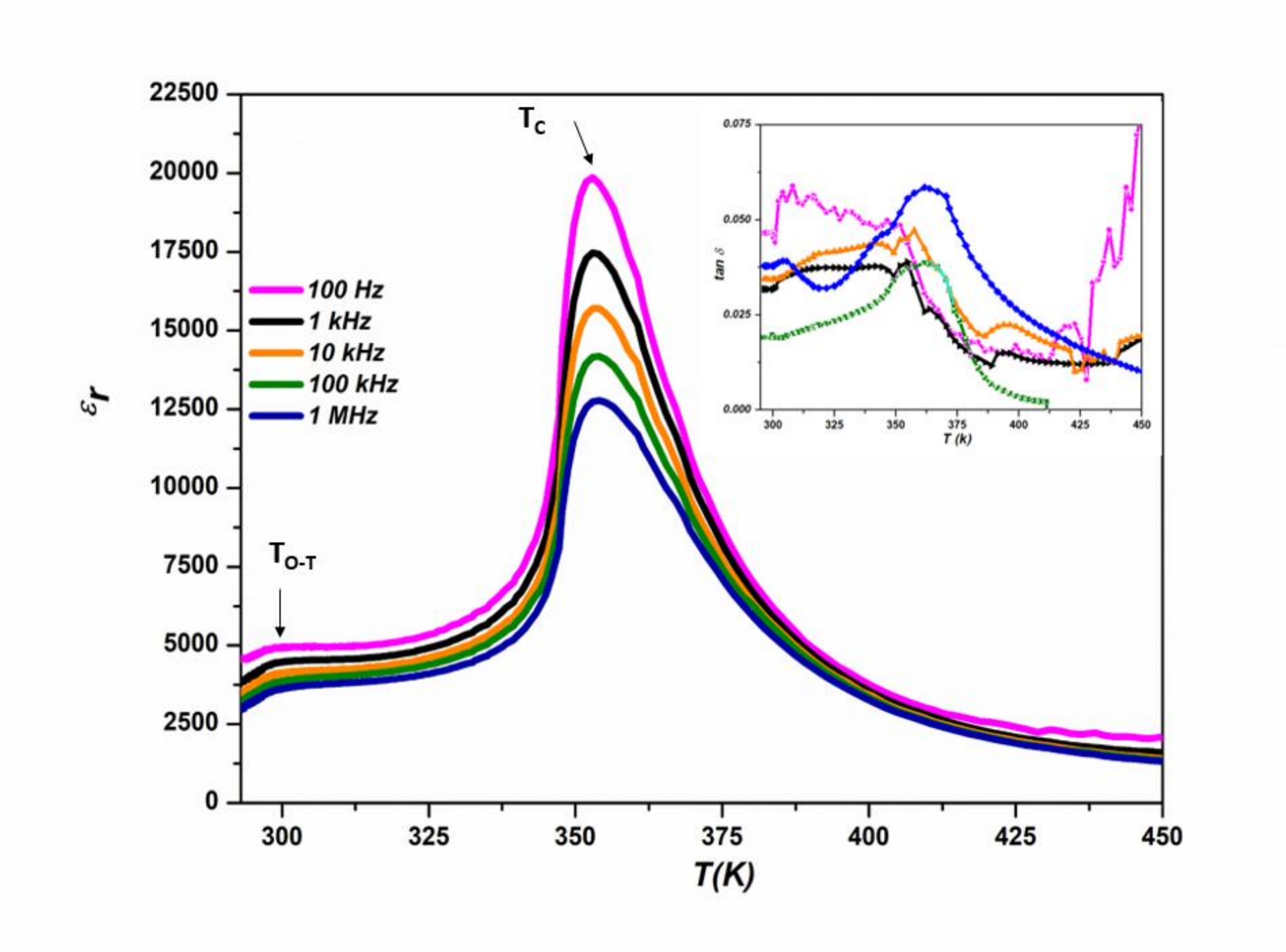

Figure 3: Temperature variation of dielectric properties at different frequencies of BCZT sample; Inset: Dielectric loss at different frequencies

## III. Ferroelectric properties

**Fig.4** show the hysteresis loops at different temperatures at an electric field of 55 kV/cm for BCZT ceramic. The measurements were carried out at different temperatures and a constant frequency of 10Hz. At 303K, the P-E loop reveals a closed and saturated loop similar to a conventional ferroelectric behavior [31]. The maximal polarization ($P_{max}$) was 25.75µC/cm², the remnant polarization ($P_r$) was 15.15 µC/cm², and the coercive field ($E_c$) was found to be 6.41 kV/cm. It is also observed that the hysteresis loops showed typical characteristics of ferroelectric to paraelectric phase transition with nonlinear to linear behavior [32]. As the temperature increases, the hysteresis loops narrow until linear behavior is achieved, and there was a continuous decrease of $P_r$, $Ec$, and $P_{max}$ due to the thermal agitation [33]. This behavior is a typical feature of paraelectric ferroelectric phase transitions. The value of $P_r$ and $P_{max}$ is upper than that of BCZT ceramic prepared by solid-state reaction ($P_{max}$ = 16.01 µC/cm²– $P_r$= 10.98 µC/cm²) [30], due to a good homogeneity in grain size and uniform grain microstructure of the samples prepared by sol-gel. Indeed, the proportion ($f$) of grains contribution to polarization for ferroelectric materials can be described by [34]:

$$f = f_0\left[1-\exp\left(\frac{-G_a d^3}{kT}\right)\right] \qquad (1)$$

Where $d$ is the grain size and $G_a$ is a constant representing the grain anisotropy energy density. Therefore, the ferroelectric properties were proportional to grain size. Thus, with the increasing grain size, the contribution of grain to polarization increases, giving rise to the enhancement of ferroelectricity [34]. Besides, a relatively lower $E_c$ was found compared to ceramic prepared by solid-state ($E_c$= 9.30 kV/cm) [30]. The large grain of ferroelectric ceramics is beneficial to a fast ferroelectric domain switching resulting in the reduced coercive electric field.

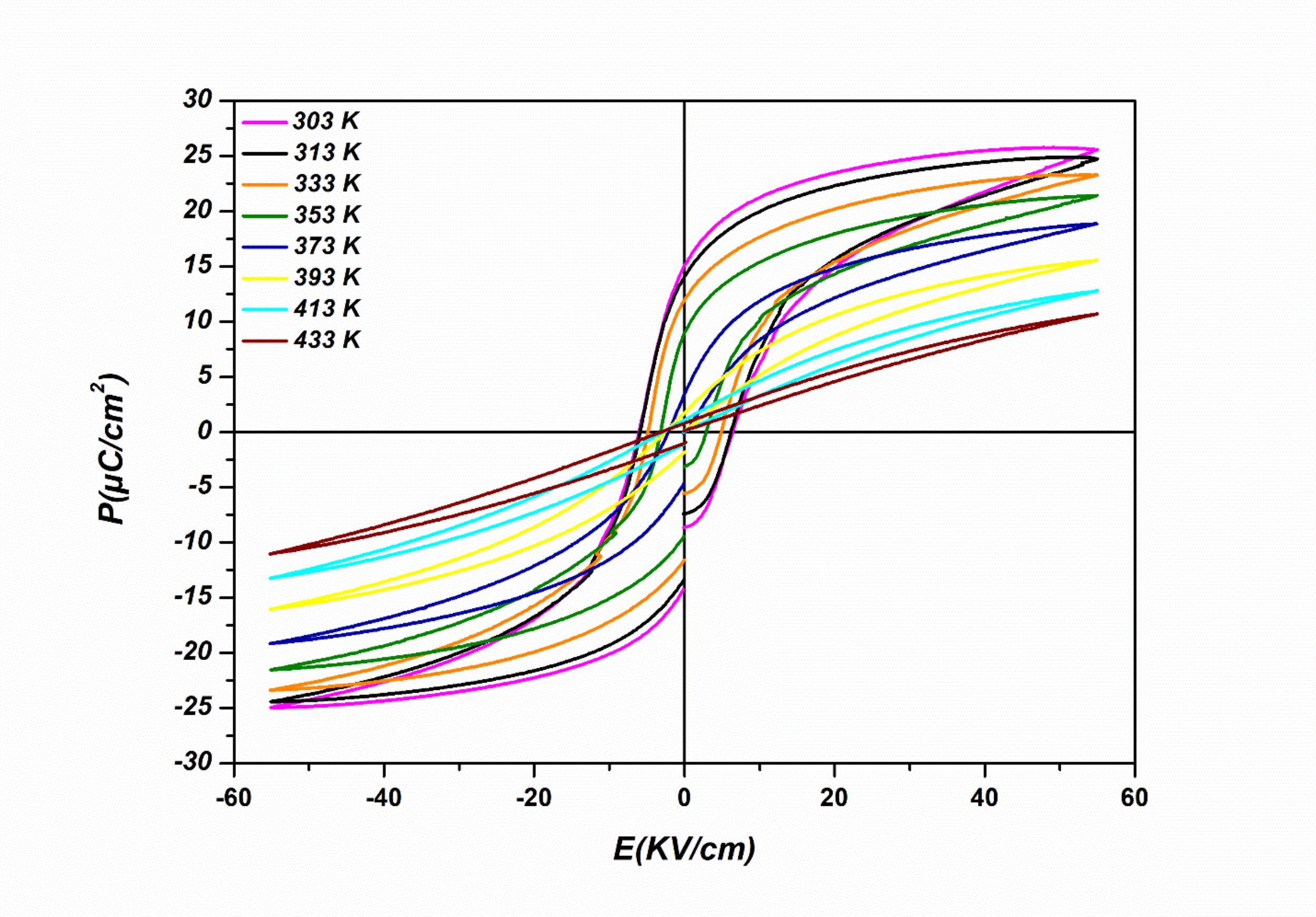

Figure 4: The hysteresis loops at different temperatures at an electric field of 55 kV/cm for BCZT ceramic

## IV. Energy storage performances

The energy storage properties of BCZT ceramic were deduced from the P–E hysteresis loops as a function of temperature. Nonlinear dielectrics possess some energy dissipation. Therefore, the total energy density ($W_{total}$), recoverable energy density ($W_{rec}$), and energy storage efficiency ($\eta$) could be calculated using the equations [35]:

$$W_{total} = \int_0^{P_{\max}} E dP \qquad (2)$$

$$W_{rec} = \int_{Pr}^{P_{max}} EdP \tag{3}$$

$$\eta = \frac{W_{rec}}{W_{total}} \times 100 \tag{4}$$

P-E loops' upper and lower branches represent the discharge and charge process, respectively. $W_{rec}$ could be determined by integrating the area between the polarization axis and the upper branch curve of the unipolar P–E hysteresis loop, $W_{rec}$ represents the energy released during the discharging process. In contrast, $W_{loss}$ produced by the domain reorientation is determined by integrating the P–E loop area and corresponds to the dissipation in dielectrics. $W_{total}$ can be calculated by adding $W_{rec}$ and $W_{loss}$ as schematically presented in **Fig. 5**, and it corresponds to total energy density during the charging process.

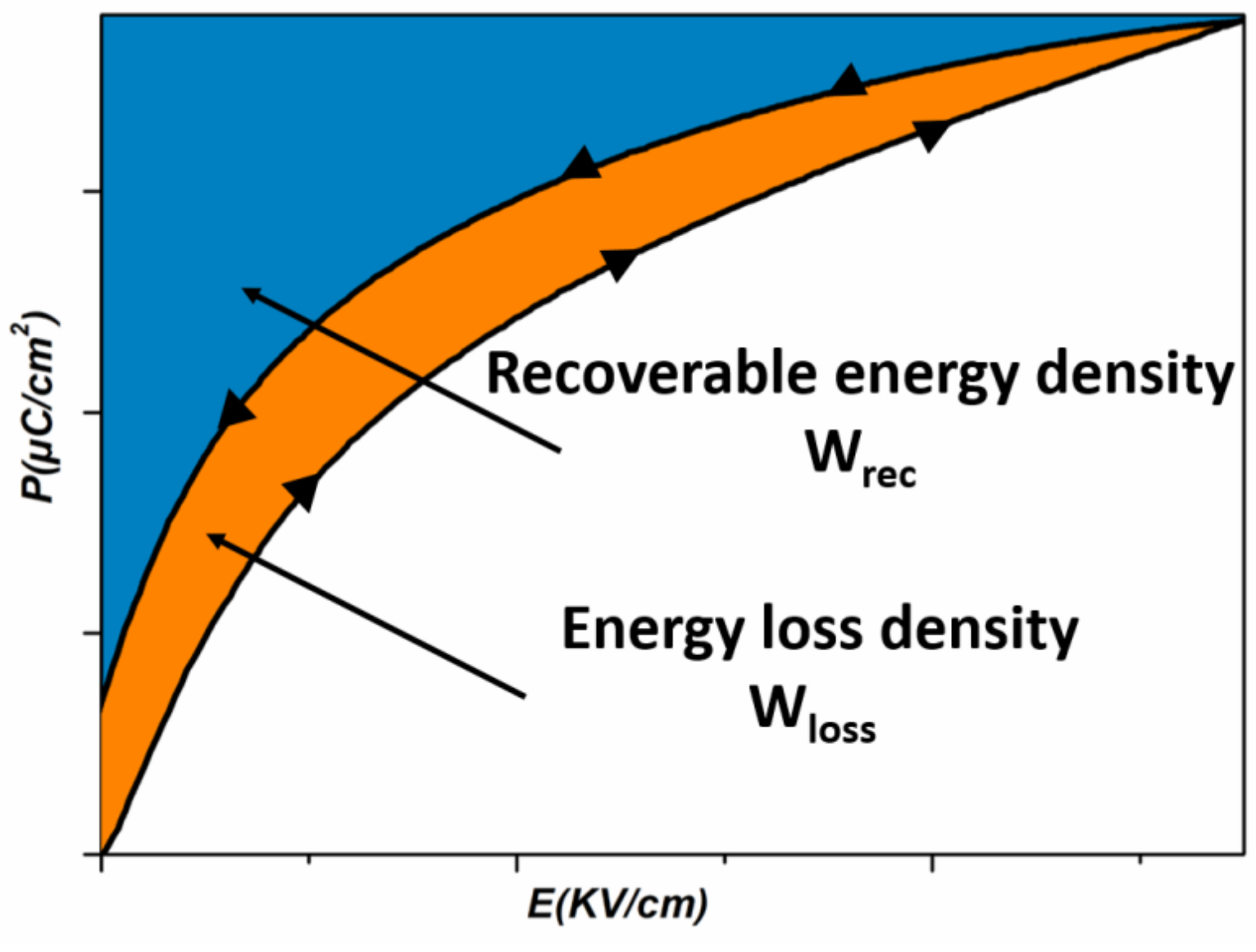

Figure 5: P–E curve with a schematic for calculating energy storage performance

**IV-1 Electric-field-dependent energy storage properties at room temperature**

Using Equations (2-4), the obtained values of total energy density $W_{total}$, energy storage density $W_{rec}$, energy loss density $W_{loss}$, and energy storage efficiency $\eta$ as a function of electric field have been illustrated in **Fig. 6**. At room temperature, $W_{total}$, $W_{rec,}$ and $W_{loss}$ display sharp linear rise with the increasing applied electric field. $W_{total}$, $W_{rec,}$ and $W_{loss}$ increased from 0.10 J/cm$^3$ to 0.55 J/cm$^3$, 0.01 J/cm$^3$ to 0.12 J/cm$^3$ and 0.08 J/cm$^3$ to 0.43 J/cm$^3$ respectively, with the increasing applied electric field from 5 kV/cm to 55 kV/cm. However, the value of $\eta$ increases steadily with increasing the electric field. From 5 kV/cm to 55 kV/cm, $\eta$ increased from 15.10 % to 24.12%. These results indicate that higher applied electric fields can induce more polarization, enhancing the energy storage properties [36].

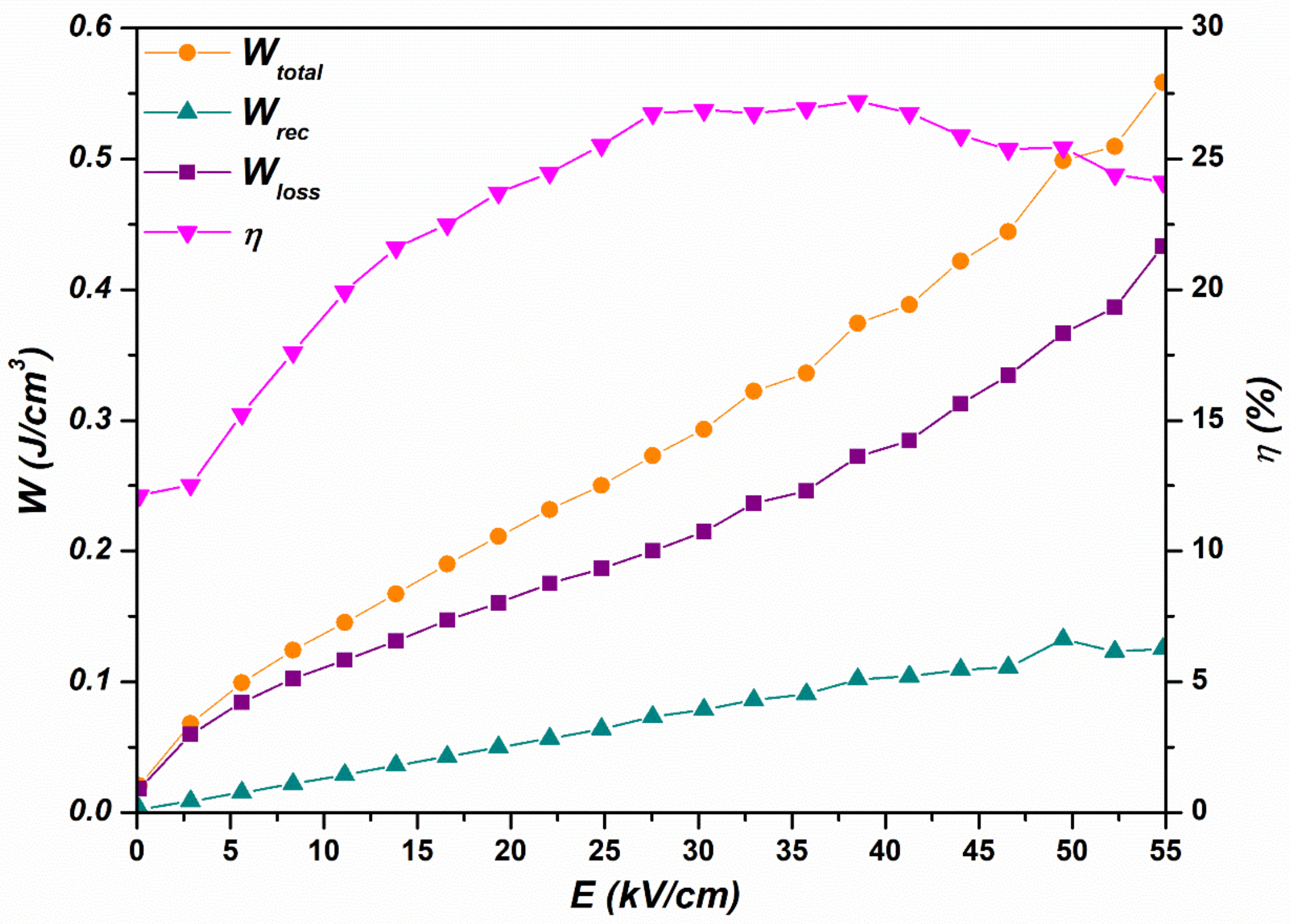

Figure 6: Total energy density ( $W_{total}$), energy storage density ($W_{rec}$), energy loss density ($W_{loss}$), and energy storage efficiency $\eta$ as a function of the electric field at room temperature for BCZT ceramic

**IV-2 Temperature-dependent energy storage properties**

Another essential aspect of energy storage applications is the high-temperature stability of dielectric ceramics. To gain insight, the temperature dependence of energy storage properties of BCZT ceramic has been studied. **Fig. 7** shows the variation of $W_{total}$, $W_{rec}$, $W_{loss}$, *and* $\eta$ for BCZT ceramic as a function of temperature. On the one hand, $W_{rec}$ increase with the increase in temperature. This behavior may be due to the enhancement in the mobility of domain walls with a rise in temperature [36]. On the other hand, a decrease in $W_{loss}$ values with increased temperature is observed. This behavior can be attributed to ease in the reorientation of domains with rising temperature [37]. As a result, energy storage efficiency is enhanced. At the maximum electric field of 55kV/cm, the maximum efficiency value is $\eta$ = 88.05 % at 324 K with $W_{rec}$= 0.24 J/cm$^3$. These results are comparable to others reported in the literature under a different electric field **(Table 2**). Using the same synthesis method and the temperature of synthesis, Mezzourh et al. [16] found a recovered energy density $W_{rec}$ of 0.02 J/cm$^3$ and storage efficiency $\eta$ of 63.65% under 7.7 kV/cm in BCZT ceramic. On the other hand, by using an electric field of 20 kV/cm, Cai et al. [38] reported a recovered energy density of 0.06 J/cm$^3$ and storage efficiency of 85.8% in BCZT ceramic elaborated by solid-state method with almost the same grain size (44.37 µm) as our samples, which is lower than our results. For BCZT synthesis by sol-gel hydrothermal and at 60 kV/cm, the obtained $W_{rec}$ was 0.36 J/cm$^3$ and $\eta$ = 67.25% with the grain size of 22.1 μm, as stated by Hanani et al. [39]. In comparison, BCZT ceramic under 55 kV/cm exhibited lower energy storage performances. Therefore, it is important to note that the energy-storage properties depend on the method of synthesis as well as the grain size. The better energy density may be due to smaller grain sizes and more uniform grain microstructure.

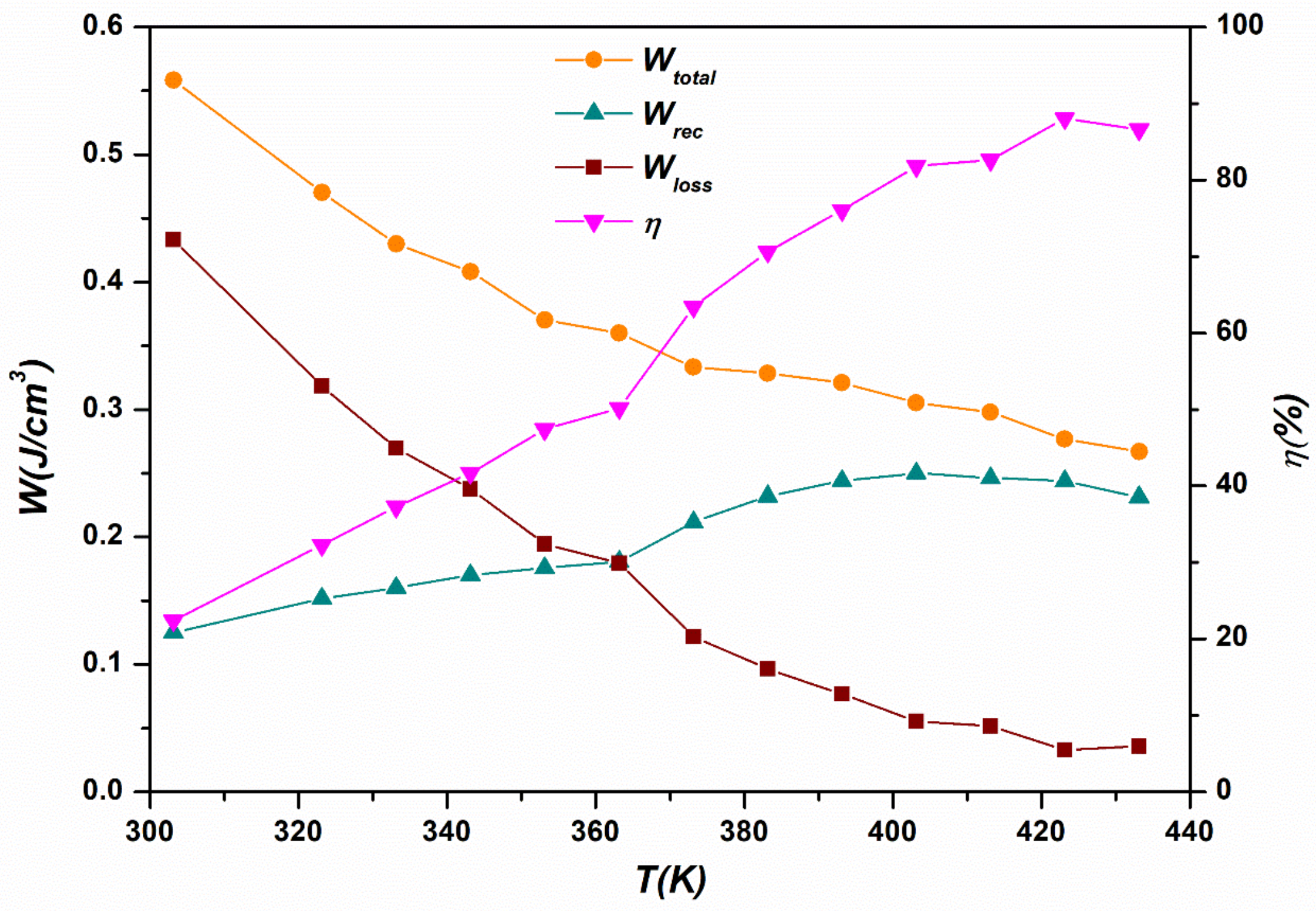

Figure 7: Total energy density ($W_{total}$), energy storage density ($W_{rec}$), energy loss density ($W_{loss}$), and energy storage efficiency ($\eta$) as a function of temperature for BCZT ceramic at 55 kV/cm

Table 2: Comparison of energy storage properties of BCZT ceramic with other lead-free materials from the literature

| Sample | Method | T of Sintering | $\varepsilon_r$' | $W_{rec}$ (J/cm$^3$) | $\eta$ (%) | $E$ (kV/cm) | $T$ (K) | Refs. |
|---|---|---|---|---|---|---|---|---|
| $Ba_{0.85}Ca_{0.15}Zr_{0.1}Ti_{0.9}O_3$ | Sol-gel | 1420 | 17658 | *0.07* | *27.43* | *27* | *RT* | This work |
| $Ba_{0.85}Ca_{0.15}Zr_{0.1}Ti_{0.9}O_3$ | Sol-gel | 1420 | 17658 | 0.12 | 24.12 | *55* | *RT* | This work |
| $Ba_{0.85}Ca_{0.15}Zr_{0.1}Ti_{0.9}O_3$ | Sol-gel | 1420 | 17658 | 0.24 | 88.05 | *55* | *324* | This work |
| $Ba_{0.85}Ca_{0.15}Zr_{0.1}Ti_{0.9}O_3$ | Sol-gel | 1420 | 16310 | 0.01 | 63.65 | 7.77 | *413* | [16] |
| $Ba_{0.85}Ca_{0.15}Zr_{0.1}Ti_{0.9}O_3$ | Solid-state | 1500 | 19277 | 0.06 | 87.4 | 20 | 373 | [38] |
| $Ba_{0.85}Ca_{0.15}Zr_{0.1}Ti_{0.9}O_3$ | Solid-state | 1350 | 16200 | 0.12 | 51.3 | 60 | 303 | [40] |
| $Ba_{0.85}Ca_{0.15}Zr_{0.1}Ti_{0.9}O_3$ | Hydrothermal | 1250 | 5000 | 0.41 | 78.6 | 60 | 380 | [39] |
| $Ba_{0.85}Ca_{0.15}Zr_{0.1}Ti_{0.9}O_3$ | Coprecipitation | 1440 | - | 0.54 | 71 | 150 | - | [41] |
| $BTSn_{0.11}$ | Solid-state | 1350 | 17390 | 0.08 | 91.04 | 25 | 333 | [42] |
| $BaZr_{0.05}Ti_{0.95}O_3$ | high energy ball milling | 1200 | 1300 | 0.10 | 56 | 30 | RT | [10] |
| $BaZr_{0.05}Ti_{0.95}O_3$ | high energy ball milling | 1200 | 1300 | 0.218 | 72 | 50 | RT | [10] |

**IV-3 Temperature-dependent energy storage density using landau phenomenological approach**

More insight into calculations of the total energy density $W_{total}$ can be obtained on the basis of the Landau-Ginzburg-Devonshire (LGD) phenomenological theory, describing the macroscopic phenomena in ferroelectric materials near the phase transition. The total energy density Wtotal can be calculated by considering the simple Landau-type free energy model [48]:

$$F = F_0 + a\,P^2 + b\,P^4 + c\,P^6 - EP \tag{5}$$

Where *a*, *b* and *c* are the temperature-dependent coefficients. In ferroelectrics, $a(T) = a_1(T - T_c)$ where $T_c$ is the curie temperature. At equilibrium, the first deviation of free energy should equal zero $(\partial F / \partial P = 0)$ [43], which thus leads to:

$$E = aP + bP^3 cP^5 \tag{6}$$

The electric field versus polarization (E-P) data for all considered temperatures can be fitted by Equation (6), which allows extracting the *a*, *b* and *c* parameters for each temperature. These coefficients are important for energy storage because they are used to express the energy density, according to the Landau model [43]; Where $P_{max}$ is the polarization at $E_{max}$.

$$W_{total} = \int_0^{P_{\max}} (aP + bP^3 + cP^5)dP = \frac{1}{2}aP_{max}{}^2 + \frac{1}{4}bP_{max}{}^4 + \frac{1}{6}cP_{max}{}^6 \tag{7}$$

**Fig. 8** compare the temperature-dependence of the estimated $W_{total}$ obtained via LGD theory with that deduced from the experimental results under 55 kV/cm in BCZT ceramic. Both methods show the same trend, $W_{total}$ decrease with increase in temperature. At room temperature, the $W_{total}$ obtained by LGD was found to be 0.51 J/cm$^3$, which is lower than that found experimentally (0.55 J/cm$^3$). However, both methods lead to the same values at the paraelectric phase transition. Note, the coefficients *a*, *b* and *c* are temperature-dependent in relaxor ferroelectrics, which could be the main reason for the difference between the results obtained by the theoretical calculation and the experimental result [44].

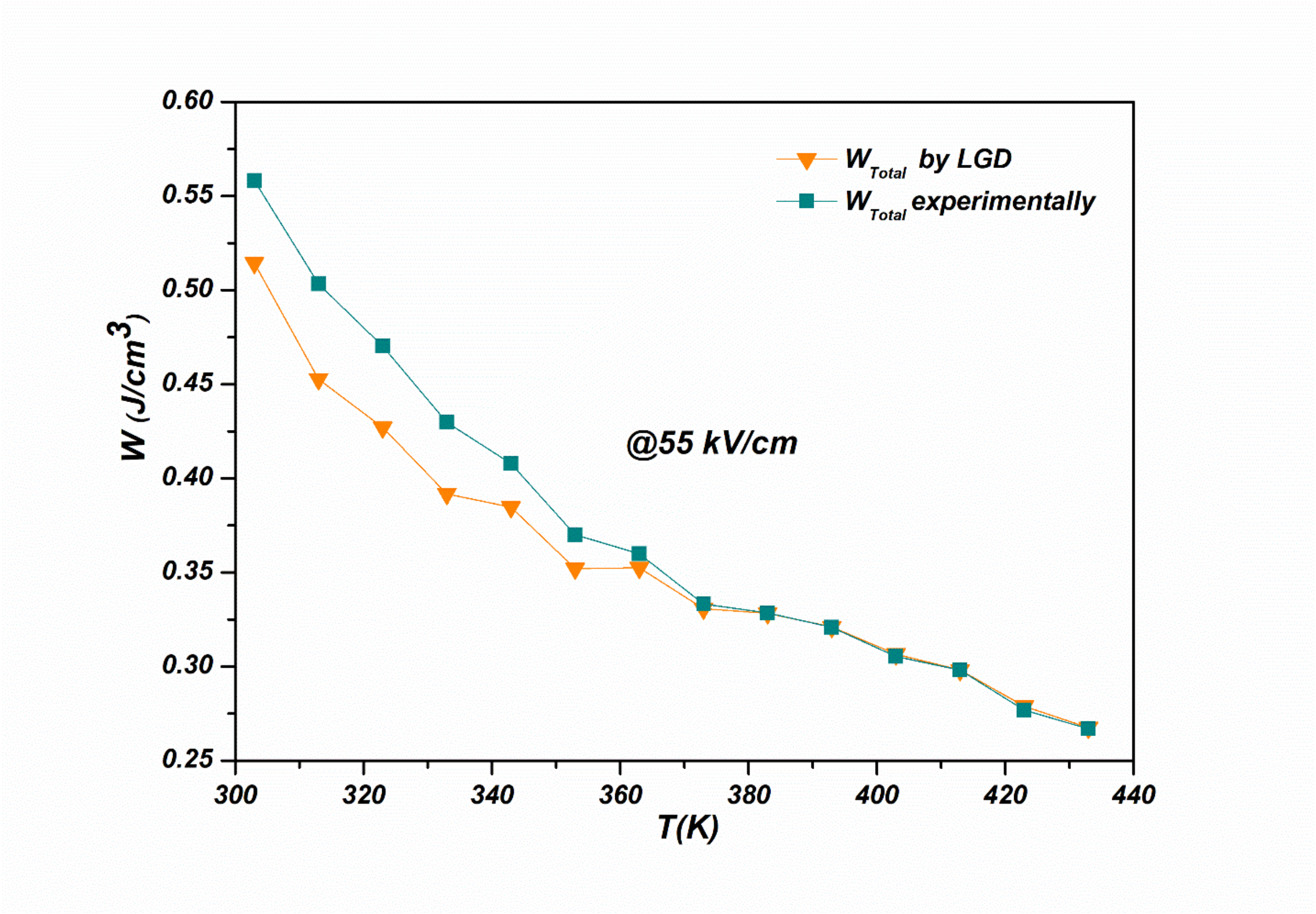

Figure 8: Comparison of the thermal evolution of $W_{total}$ obtained experimentally and using Landau-Ginzburg-Devonshire (LGD) theory at 55 kV/cm for BCZT ceramic

## V. Indirect electrocaloric measurements

The electrocaloric effect (ECE) is universally known as the reversible entropy and temperature changes that occur when an electric field is applied or removed [45]. An indirect method based on Maxwell's relation between the entropy ($S$) and the dielectric polarization ($P$) is used to estimate the adiabatic temperature change $\Delta T$ [46]:

$$\left(\frac{\partial P}{\partial T}\right)_E = \left(\frac{\partial S}{\partial E}\right)_T \quad (8)$$

$\Delta T$ under an applied electric field is calculated by the following equation:

$$\Delta T = -\frac{1}{\rho}\int_{E_1}^{E_2} \frac{T}{C_P}\left(\frac{\partial P}{\partial T}\right)_E dE \tag{9}$$

Where $P$ is polarization, $\rho$ is the density of the sample, and $C_p$ denotes the specific heat capacity of the material. $E_1$ and $E_2$ are initial and final applied electric fields, respectively. In the present case, $E_1$ is taken as zero, and $E_2$ relates to the saturation field value. The value of polarization gradient $\left(\partial P/\partial T\right)_E$ was found from the derivation of $P$ versus $T$ data [37].

Temperature change alone does not provide sufficient data to determine the global EC performance. Thus, the isothermal entropy change ($\Delta S$) is an important parameter to evaluate the relative heat extraction capacity of the material. $\Delta S$ is calculated by the following equation:

$$\Delta S = -\frac{1}{\rho}\int_{E_1}^{E_2} \left(\frac{\partial P}{\partial T}\right)_E dE \tag{10}$$

**Fig. 9a-b** shows the variation of $\Delta T$ and $\Delta S$ as a function of temperature at different applied electric fields for BCZT ceramic. It has been observed that $\Delta T$ and $\Delta S$ increase with the increase in the electric field. This behavior is as per equations (9) and (10). $\Delta T$ and $\Delta S$ curves, which corresponds to a fixed applied electric field, exhibit a maximum around $T_c$ due to the diffusion behavior of our sample near the ferroelectric-to-paraelectric phase transition and multiphase coexistence [47]. From equations (9) and (10), it is evident that materials having a more significant polarization gradient as well as a higher saturation field ($E_2$) will show larger $\Delta T$. At $\Delta E_{max}$ of 55 kV/cm, $\Delta T_{max}$ was 2.32K, and $\Delta S_{max}$ was 1.98 J/Kg/K for BCZT ceramic with thickness of 400µm. The maximum peak is near transition temperature for lower electric fields. However, the peak position of $\Delta T_{max}$ and $\Delta S_{max}$ moves gradually toward the higher temperature with increasing the intensity of the applied electric field. This peak shift relates to the diffused nature of phase transition, which might be due to a change in domain configuration or field-induced transition on the application of higher electric fields [36].

However, comprehensive conclusions cannot be drawn alone with adiabatic temperature change ($\Delta T$) values about the overall caloric performance of a ferroelectric material. Thus, the electrocaloric responsivity ($\Delta T/\Delta E$), i.e., the ratio of adiabatic temperature change and the field change, is an important parameter to evaluate the ECE behavior of the material [48]. At 384K, the maximum electrocaloric responsivity $\zeta_{max}$ is 0.42K mm/kV at 55 kV/cm for BCZT ceramic. The EC effect compared for different materials, electric fields, and thickness are gathered in (**Table 3**). We can observe that the same material with the same composition shows different ECE. Certainly, for the BCZT with 1mm of thickness and synthesized by solid-state, $\zeta_{max}$ reached 0.23 Kmm/kV at 25 kV/cm [30], which is lower than our results. On the other hand, with the same ceramic's thickness of 1 mm, the ECE of BCZT at 27 kV/cm was higher than many other lead-free ceramics. Indeed, Patel et al. [32] elaborated the $(Ba_{0.85}Ca_{0.15})(Zr_{0.1}Ti_{0.88}Sn_{0.02})O_3$ ceramics by solid oxide reaction, revealing that the ECE response reached high values of 0.84 K and 0.26 Kmm/kV for $\Delta T_{max}$ and $\zeta_{max}$, respectively, under 32 kV/cm at 347 K . As well as $(Ba_{0.85}Sr_{0.15})(Zr_{0.1}Ti_{0.9})O_3$ ceramics fabricated by conventional solid-state reaction route, which $\Delta T_{max}$ = 0.5K and $\zeta_{max}$ = 0.16 Kmm/kV at 347 K at 30 kV/cm, as stated by Patel et al. [49]. At an electric field of 55 kV/cm, BCZT exhibited better ECE properties compared to those found in $0.98(BaTiO_3)$–$0.02\ Bi(Mg_{0.5}\ Ti_{0.5})O_3$ ceramics with 2 mm of a thickness ($\Delta T_{max}$ = 1.21K and $\zeta_{max}$ = 0.22 Kmm/kV at 416 K) by Li et al. [50]. The synthesis conditions (such as calcination and sintering), chemical doping, grain form, the number of coexisting phases, the ceramic thickness, and the applied external electric field are all factors that contribute to these differences.

It is crucial to remember that the frequency and temperature interval of measurements can affect the ECE values derived from indirect measurements utilizing PE hysteresis loops. It can also be affected by $C_p$, which is supposed to be constant. For that reason, refrigeration capacity (*RC*) is evaluated to look at the overall material's performance. *RC* is calculated by the following equation:

$$RC = \Delta T \times \Delta S \tag{11}$$

**Fig. 10** shows the variation of *RC* as a function of temperature at 55 kV/cm for BCZT ceramic. *RC* increases with increasing electric field. *RC* shows a maximum around $T_c$ due to phase transition, as discussed before. At 55 kV/cm, $RC_{max}$ was estimated to be 4.59 J/kg for BCZT with thickness of 400 µm. The obtained value was higher than many other lead-free ceramics [33], [50].

Another important measure in evaluating the EC performance is the coefficient of performance (*COP*) as calculated by the following equation, where *Q* is the isothermal heat.

$$COP = \left| \frac{Q}{W_{total}} \right| = \left| \frac{T \Delta S}{W_{total}} \right| \tag{12}$$

**Fig. 10** displays also the variation of *COP* as a function of temperature at 55 kV/cm for BCZT ceramic. The *COP* reveals the same behaviors as $\Delta T$, $\Delta S$ and *RC*. Around $T_c$, *COP* exhibits a maximum. At 55 kV/cm, $COP_{max}$ was 12.38 for BCZT ceramic. The obtained value is higher than that obtained by other literature [39], [51]. Nanocrystalline ceramics of $[(k_{0.5}na_{0.5})nbo_3]_{0.97}$-$[lisbo_3]_{0.03}$ shows a *COP* of 8.14 under an electric field of 40 kV/cm as reported by Kumar et al. [52]. Whereas, Hao et al.[53] report a *COP* of 2.9 in $Pb_{0.97}La_{0.02}Zr_{0.87}Sn_{0.08}Ti_{0.07}O_3$ antiferroelectric thick film. Attending a strong electric field by reducing the thickness of ceramics can improve the EC parameters. The obtained high values of $\Delta T$, $\Delta S$, $\zeta$, *RC* and *COP* make BCZT ceramic with low thickness a promising candidate for eco-friendly electrocaloric applications.

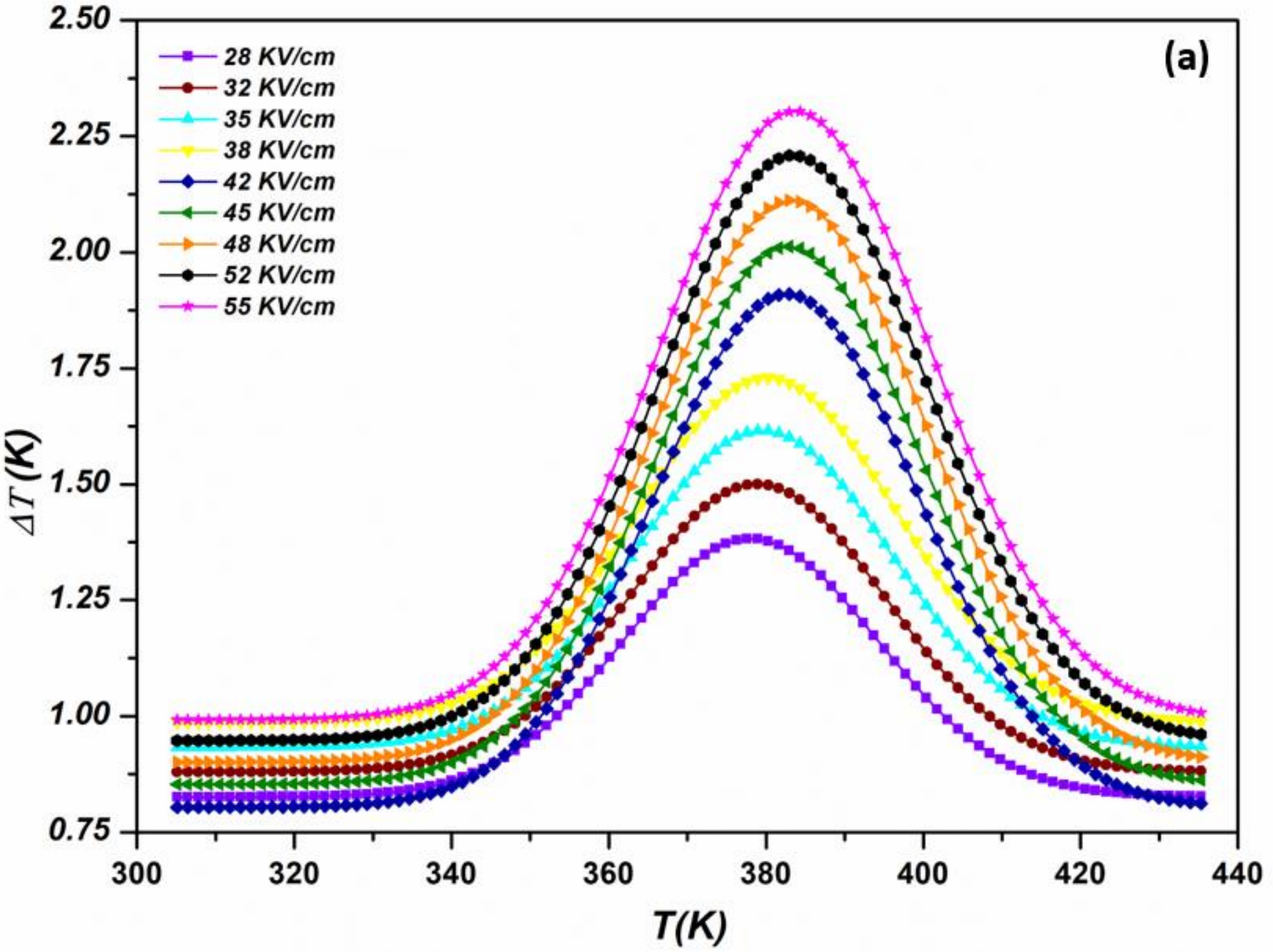

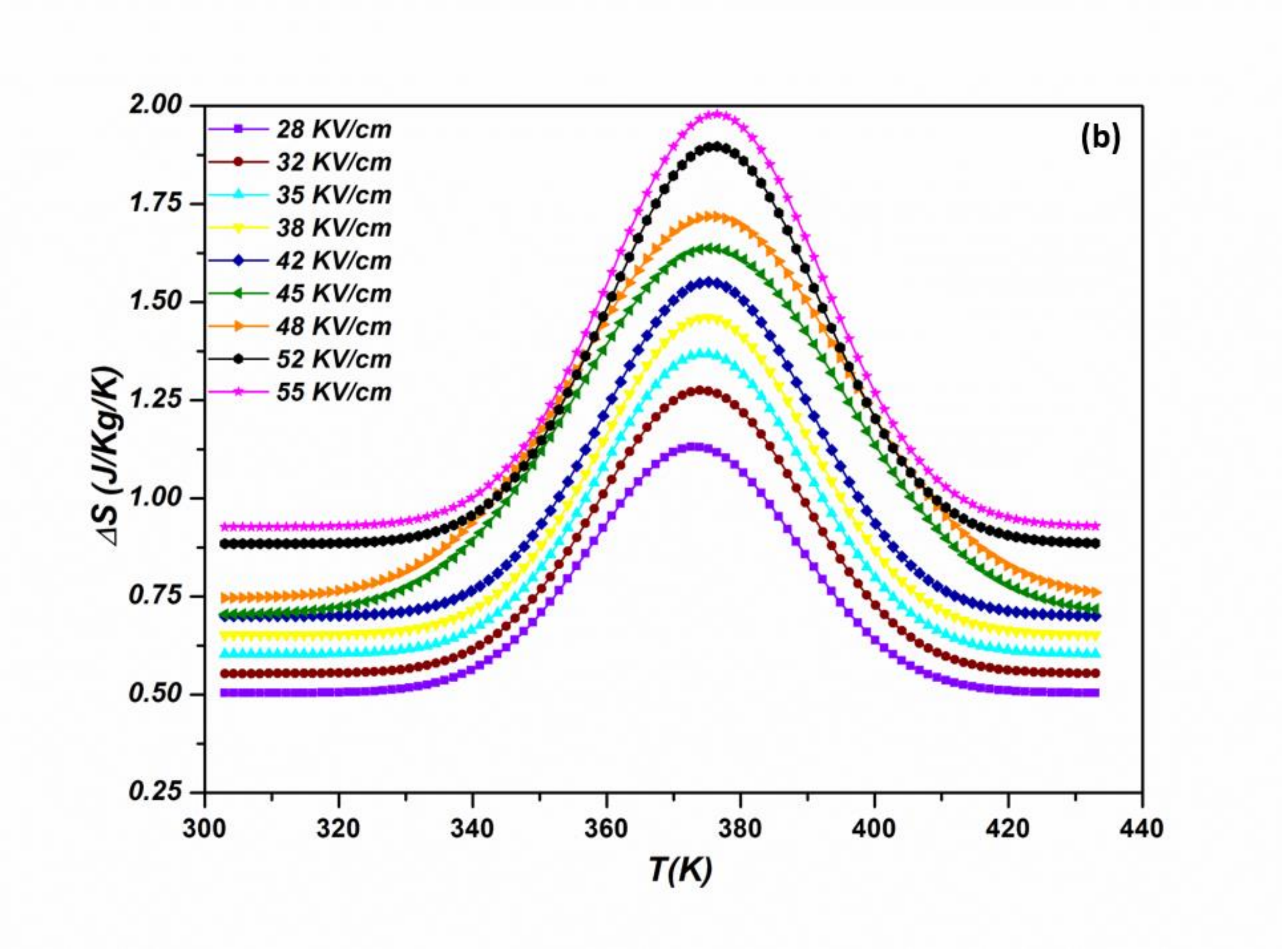

Figure 9: (a) Temperature profiles of the electrocaloric temperature change ($\Delta T$) at various applied electric fields; (b) Temperature profiles of the isothermal entropy change ($\Delta S$) at various applied electric fields

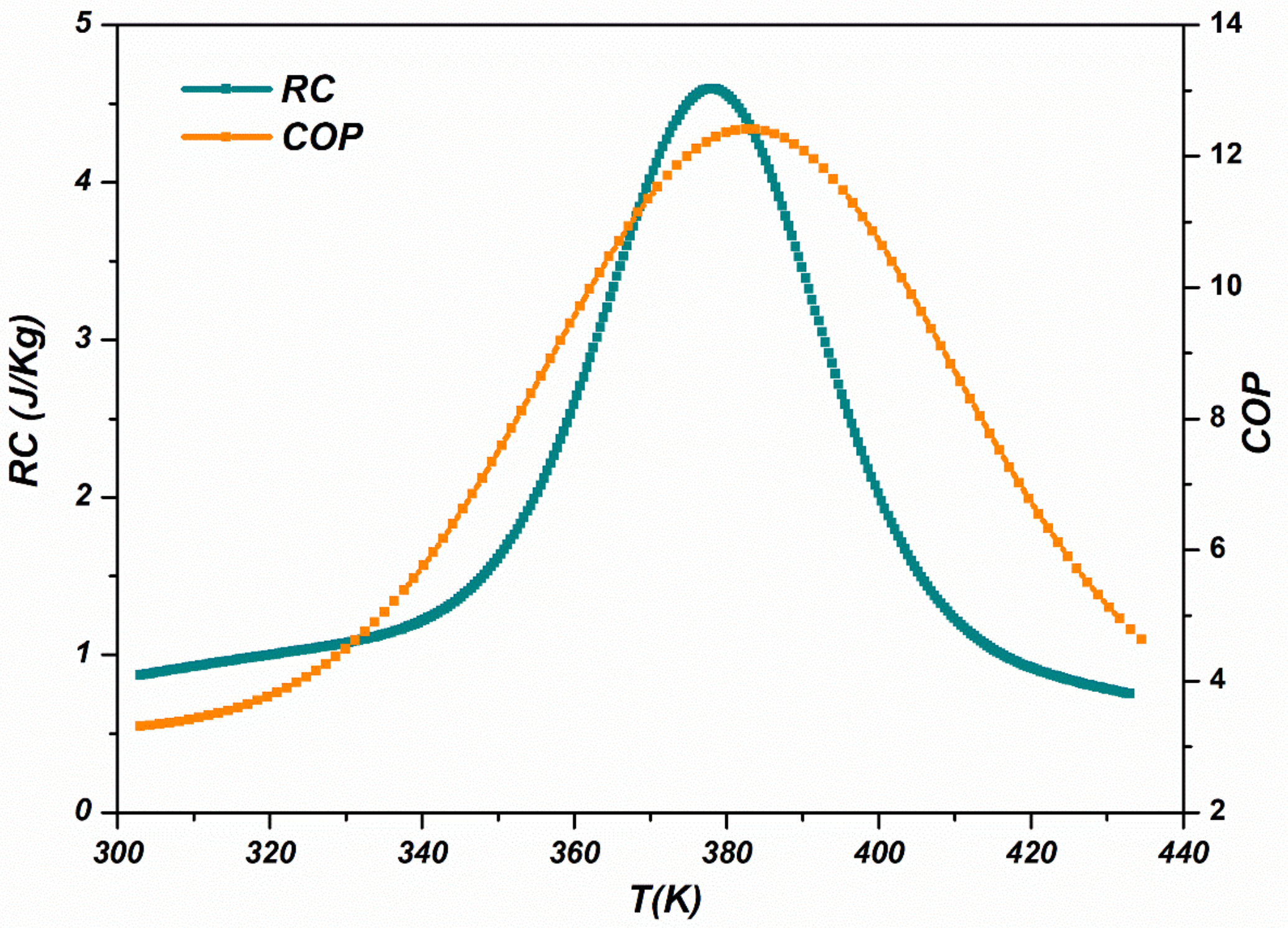

Figure 10: Temperature profiles of refrigeration capacity (*RC*) and the coefficient of performance (*COP*) at 55 kV/cm for BCZT ceramic

The temperature change, $\Delta T$, can also be calculated by using Landau-Ginzburg-Devonshire phenomenological theory. The free energy can be definite as equation (5) in terms of electric polarization. The entropy change is obtained from equation (13) [54].

$$\Delta S = -\left(\frac{\partial F}{\partial T}\right)_E = -(\frac{1}{2}a_1(T+\Delta T)P^2(E,T+\Delta T) - \frac{1}{2}a_1(T)P^2(0,T)) \tag{13}$$

The EC response can be given by using $\Delta T = -\left(T\,{\Delta S}/{C_p}\right)$ [54] as follows:

$$\Delta T = -\left(\frac{T\Delta S}{C_p}\right) = -\frac{T}{2C_p}(a_1(T+\Delta T)P^2(E,T+\Delta T) - a_1(T)P^2(0,T)) \tag{14}$$

where $C_p$ is the specific heat capacity of the material at given field E. Since $\Delta T \ll T$, equation 14 can be simplified to:

$$\Delta T = -\frac{T}{2C_p}\left(a_1(T)P^2(E,T) - a_1(T)P^2(0,T)\right) \quad (15)$$

**Fig. 11** compare the temperature-dependence of the estimated $\Delta T$ obtained via LGD theory with that deduced from the experimental results under 55 kV/cm in BCZT ceramic. Both methods show the same behavior, $\Delta T$ increase with temperature increasing, reach a maximum around $T_c$, and then decrease. This comportment is due to the diffusion behavior of our sample near the ferroelectric-to-paraelectric phase transition [47] as discussed before. At room temperature, the $\Delta T$ obtained by LGD was found to be 2,33 K, which is lower than that found experimentally (2,11 K). However, the same values at the paraelectric phase transition were observed for both methods. The difference between the results obtained by the theoretical calculation and the experimental result may be due to the coefficients *a*, *b* and *c*, which are temperature-dependent in relaxor ferroelectrics [54].

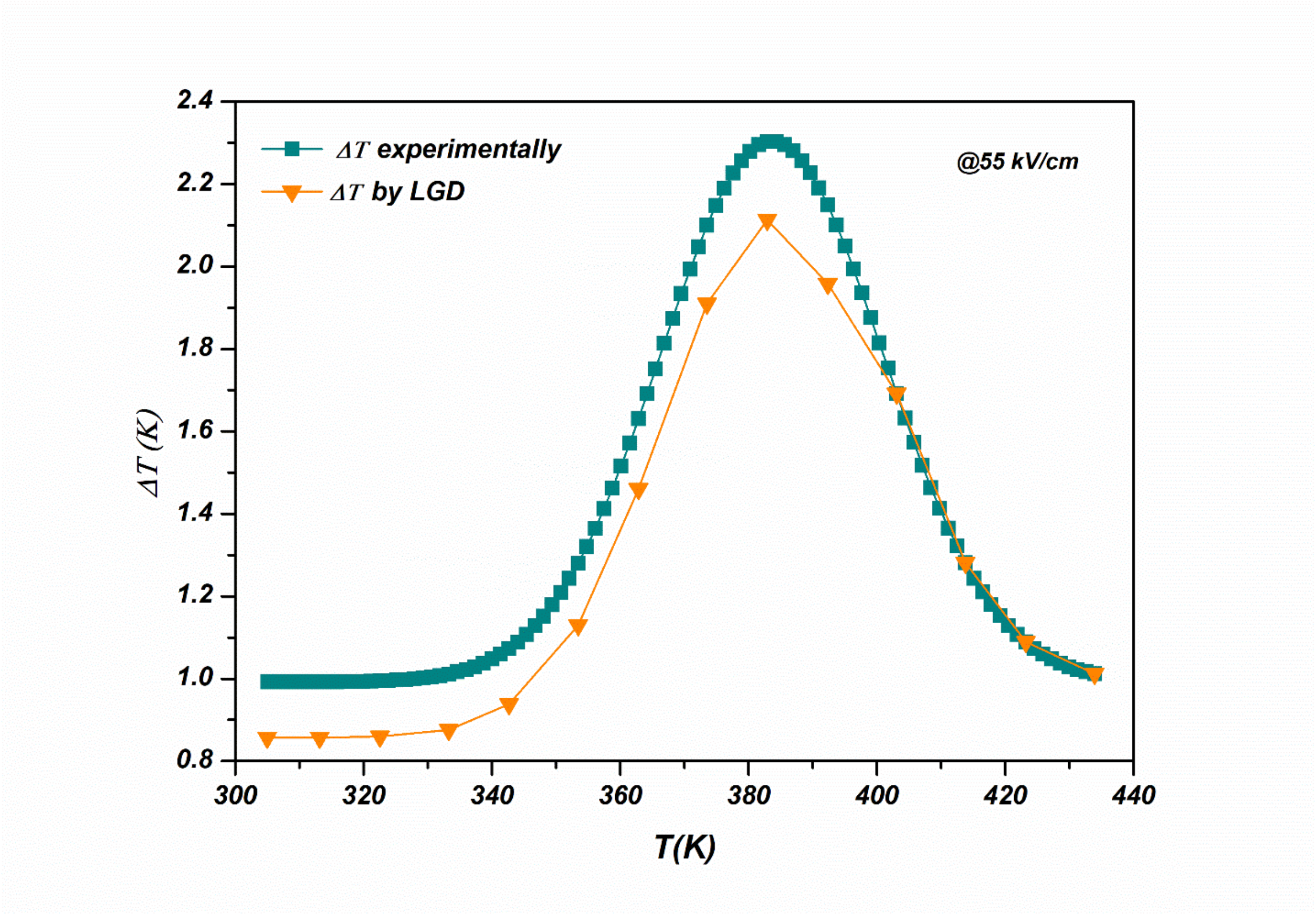

Figure11: Comparison of the thermal evolution of ΔT obtained by Landau-Ginzburg-Devonshire (LGD) theory and experimentally at 55 kV/cm for BCZT ceramic

Table 3: Comparison of the ECE properties of BCZT ceramic with other lead-free materials from the literature

| Materials | T (K) | $\Delta T$ (K) | $\Delta E$ (kV/cm) | $\zeta max$ (Kmm/kV) | Thickness (mm) | *RC* (J/Kg) | *COP* | Refs. |
|---|---|---|---|---|---|---|---|---|
| $Ba_{0.85}Ca_{0.15}Zr_{0.1}Ti_{0.9}O_3$ | 384 | 2.32 | 55 | 0.42 | 0.4 | 4.59 | 12.38 | This work |
| $Ba_{0.85}Ca_{0.15}Zr_{0.1}Ti_{0.9}O_3$ | 367 | 1.47 | 60 | 0.24 | 0.34 | 2.38 | - | [39] |
| $(Ba_{0.85}Ca_{0.15})(Zr_{0.1}Ti_{0.88}Sn_{0.02})O_3$ | 353 | 0.84 | 32 | 0.26 | 1 | - | - | [32] |
| $(Ba_{0.85}Ca_{0.075}Sr_{0.075})(Zr_{0.1}Ti_{0.88}Sn_{0.02})O_3$ | 305 | 1.5 | 33 | 0.455 | 1 | 2.75 | - | [33] |
| $(Ba_{0.85}Sr_{0.15})(Zr_{0.1}Ti_{0.9})O_3$ | 347 | 0.5 | 30 | 0.16 | 1 | - | - | [49] |
| $0.98(BaTiO_3)$–$0.02\ Bi(Mg_{0.5}\ Ti_{0.5})O_3$ | 416 | 1.21 | 55 | 0.22 | 2 | 1.47 | - | [50] |
| $Ba_{0.85}Ca_{0.15}Zr_{0.1}Ti_{0.9}O_3$ | 373 | 0.57 | 25 | 0.23 | 1 | 0.35 | 11 | [51] |
| $0.45(Ba_{0.2}Ti_{0.}8O_3)0.55(Ba_{0.7}Ca_{0.3}TiO_3)$ | 404 | 0.46 | 12 | 0.38 | 0.75 | - | - | [55] |
| $Ba_{0.4}Sr_{0.6}Nb_2O_6$ | 407 | 0.32 | 60 | 0.05 | 0.6 | - | - | [56] |
| $0.9(K_{0.5}N_{a0.5})NbO_3$ – $0.1(SrZrO_3)$ | 367 | 1.19 | 35 | 0.34 | 1 | - | 0.62 | [57] |
| $(Ba_{0.95}Ca_{0.05})(Ti_{0.89}Sn_{0.11})O_3$ | 335 | 0.80 | 30 | 0.268 | - | - | 8.80 | [58] |

## Conclusion

$Ba_{0.85}Ca_{0.15}Zr_{0.1}Ti_{0.9}O_3$ ceramic sintered at 1420 °C for 4h was successfully prepared by the sol-gel method. The XRD showed the coexistence of two tetragonal and orthorhombic phases at room temperature. The SEM revealed a large grain size of~ 45 µm. To achieve high-energy storage and large electrocaloric effect in bulk ceramics, the ceramic thickness was reduced. The BCZT ceramic with a thickness of 400µm showed both interesting energy storage properties and a significant electrocaloric effect. The recoverable energy density $W_{rec}$ was 0.24 J/cm$^3$, and $W_{total}$ was 0.27 J/cm$^3$ with an efficiency coefficient of ~ 88% at 423 K under 55 kV/cm electric field. On the other hand, BCZT show an outstanding temperature change $\Delta T$= 2.32K and significant electrocaloric responsivity $\zeta$ = 0.421 Kmm/kV, high refrigeration capacity *RC*= 4.59 J/kg and interesting coefficient of performance *COP* =12.38. The total energy density $W_{total}$ and the temperature change $\Delta T$ were also calculated by exploiting the Landau-Ginzburg-Devonshire (LGD) theory. The theoretical result matched the experimental findings. These characteristics suggest that this eco-friendly BCZT ceramic could be a prospective candidate for dielectric energy storage and electrocaloric applications.